\documentstyle[ijp,twoside,multicol,epsf,color]{ijptex}

\def\prl{{\it Phys. Rev. Lett. }}
\def\pre{{\it Phys. Rev. E }}
\def\euro{{\it Europhys. Lett. }}
\def\pla{{\it Phys. Lett. A }}
\def\ijmp{{\it Int. J. Mod. Phys. B }}
\def\pjp{{\it Pramana J. Phys. }}
\def\ajp{{\it Am. J. Phys. }}
\begin{document}
\title{\LARGE \bf{AC driven thermal ratchets.}}
\author{Debasis Dan, A. M. Jayannavar}
\author{\it{Institute of Physics, Sachivalaya Marg, Bhubaneswar 751005,
 India}}
\author{Mangal C. Mahato}
\author{\it{Department of Physics, Guru Ghasidas University, Bilaspur 4
95009, India}}
\email{dan@iopb.res.in}
\maketitle
\begin{abstract}

We consider the motion of a overdamped Brownian particle in periodic
asymmetric potential  with space 
dependent friction coefficient. In the presence of external time
periodic forcing, the system shows multiple current reversals on
varying the amplitude of the external forcing and the temperature of the thermal    
bath. In the adiabatic regime we find a single reversal of current 
as a function of noise strength which can  only be accounted due
to the presence of space dependent friction coefficient. For very
large forcing term, the current does not go to zero, instead it 
asymptotically tends to a limiting value depending on the phase shift
between the potential and the friction. This fact plays an important
role in obtaining  multiple current reversals.
\end{abstract} 
\thispagestyle{empty}
\pagestyle{myheadings}
\markboth{{\it D. Dan, M. C. Mahato and A. M. Jayannavar}}{
{\it AC driven thermal ratchets }}
\begin{multicols}{2}

\section{Introduction}
          Rectification of noise leading to unidirectional motion in
ratchet systems has been an  
active field of research over the last decade. In these systems
directed Brownian motion of particles is induced by nonequilibrium noise in the
absence of any net macroscopic forces and potential gradients. Several models have been proposed to explain
this transport mechanism under various nonequilibrium situation, like
(a) Flashing ratchets ~\cite{Flash,Flash2}, wherein the particles experience a fluctuating
energy profile, (b) Rocking ratchets~\cite{Rock,Main}, where the Brownian particles
experiences a spatially uniform, time periodic uniform force,
$F(t+\tau) = F(t)$, (c) Diffusion ratchets~\cite{Diffu}, wherein particles are
driven by a time periodic diffusion coefficient $ D(t) = D(t+\tau) $
and (d) Correlation ratchets~\cite{Corr}, where particles are driven by spatially
uniform but temporally correlated force $\zeta(t)$ of zero
average. In these systems to get a unidirectional current either
spatially asymmetric periodic potentials or time asymmetric external
forces are necessary. The recent burst of work on ratchet systems is
motivated in  part to explain the
unidirectional transport of molecular motors in biological systems and 
a new prospect of novel techniques for separation or
segregation of particles of macro-meter size. This technique for particle separation is
based on the property of current reversal in ratchet by
varying the strength of thermal noise, amplitude of the forcing, size
of the particle or any other relevant variable in the problem. 

                In the area of Brownian rectifiers or ratchets the
study of current reversal has become a subject by itself. Bartussek     
\textit{et. al.}~\cite{Main} showed the occurrence of current reversals in
rocked thermal ratchet as a function of amplitude of rocking force as well as
the temperature of thermal bath. Multiple current reversals
have also been shown in the deterministic limit of these ratchets when
the inertial term is taken into account. However, 
these multiple current reversals in inertial ratchets are not robust
in the presence of noise. Beside rocking ratchets, 
current reversals have also been observed in flashing
ratchets~\cite{Bier,Flash2}. We in this work report that
that \textit{multiple reversals} can be achieved even in  rocked
\textit{overdamped} ratchet
in the presence of space dependent mobility, as a function of the
noise strength and amplitude of the rocking force. 
In the deterministic overdamped case we get current reversal as a function
of the amplitude of rocking force.
It is to be noted that systems with space dependent friction are not uncommon.
Brownian motion in confined geometries show space dependent friction~\cite{Conf}. It is believed that molecular motor
proteins move
close along the periodic structure of microtubules and will therefore
experience a position dependent mobility~\cite{Spc_mob}. Frictional 
inhomogeneities are common in super lattice structures and Josephson
junctions etc.

\section{Model}
      We consider an overdamped Brownian particle moving in a
asymmetric potential V(x) with space dependent friction coefficient
$\eta$(x) under the influence of external force field $F$(t) at
temperature $T$.
Throughout our analysis we take the ratchet potential
V(x) = $-\frac{1}{2 \pi} (\sin (2\pi x) + \frac{\mu}{4} \sin (4 \pi
x))$. Here $\mu$ is the asymmetry parameter with values taken in the
range $0 <\mu <1 $,  friction coefficient $\eta (x) = \eta_{0}(1-\lambda \sin
(2\pi x + \phi))$, $|\lambda| < 1$. $\phi$ determines the relative
phase shift between  friction coefficient and potential .
The forcing term is
taken to be F(t) = A$\sin (\omega t + \theta)$, ($\omega = \frac{2
  \pi}{\tau}$, where $\tau$ is period of force) . Without any
loss of  generality $\theta$ is taken to be zero.
The Langevin equation for this system in the overdamped limit
is given by~\cite{IntMod,Pram}
\begin{equation}
 \dot{x} =  -\frac{(V'(x) - F(t))}{\eta (x)} - k_{B}T \frac{\eta '(x)}{(\eta
   (x))^{2}} + \sqrt{\frac{k_{B}T}{\eta (x)}}\xi (t) ,
 \label{Langvn}
\end{equation}
where $\xi(t)$ is Gaussian thermal noise with correlation
$<\xi (t) \xi (t')> = 2 \delta (t-t')$
  The Fokker Planck Equation ( FPE)~\cite{IntMod} corresponding to
  Eqn.~(\ref{Langvn}) is given by

\begin{equation}
 \frac{\partial P(x,t)}{\partial t} = \frac{\partial}{\partial x}
 \frac{1}{\eta (x)} [k_{B}T \frac{\partial}{\partial x} + (V'(x) -
 F(t))] P(x,t) .
 \label{FPE}
\end{equation}

Here $P(x,t)$ is the probability density at position $x$ at time
$t$. Equation~(\ref{FPE}) can be recasted to a continuity equation
$\frac{\partial P(x,t)}{\partial t} = - \frac{\partial
  J(x,t)}{\partial x}$, where

\begin{equation}
  J(x,t) = -\frac{1}{\eta(x)} [(V'(x)-F(t)) + k_{B}T
  \frac{\partial}{\partial x}]P(x,t),
 \label{Current}
\end{equation}
is the probability current density. Since the potential and the driving force
have spatial and temporal periodicity respectively, therefore $J(x,t)
= J(x + 1,t + \tau)$, ~\cite{Main,Risk}. The net current $j$
in the system is given by
$ j = \lim_{t \rightarrow \infty} \frac{1}{\tau} \int_{t}^{t+\tau} dt
 \int_{0}^{1} J(x,t) dx $ .
It should be noted that for symmetric potential and $\lambda = 0$, $j
= 0$. Rectification of current is possible if the potential is either
asymmetric or $\lambda \neq 0$ with $\phi \neq 0, \pi$
~\cite{Dan3}. It is important to note that in these systems
rectification is due to a symmetry breaking as the potential is
asymmetric and $\lambda = 0$. For the case when the potential is
symmetric and $\lambda \neq 0$, the symmetry breaking results from the 
dynamics of the system. It is the space dependent mobility that breaks 
the symmetry in this nonequilibrium problem and the magnitude of the
symmetry breaking is related to the phase shift between the friction
and potential profile. $j$ is
independent of the initial phase $\theta$ of the driving force. We
solve the FPE numerically by method of finite difference and calculate 
the current $j$. All the physical quantities such as $j, T, A, \omega$ are in dimensionless units~\cite{Risk,Dan3,PLA}.
\section{Results and Discussions}

\begin{figure}
\protect\centerline{\epsfysize=3.0in \epsfbox{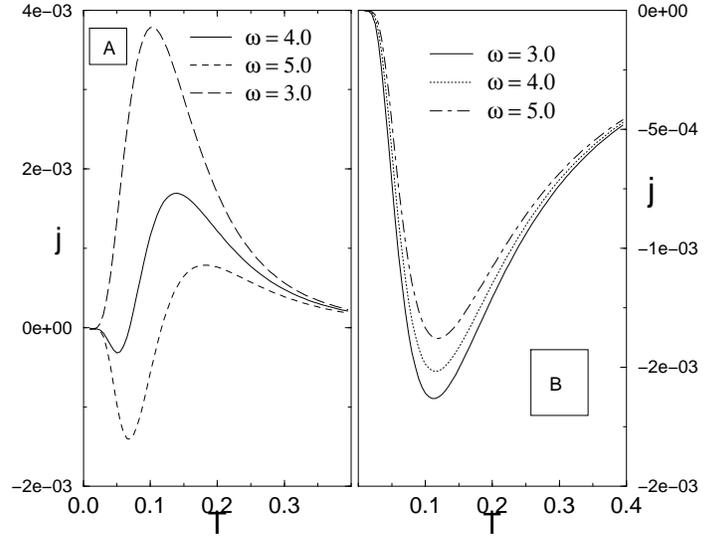}}
\caption{Mean current j vs temperature T for (a) $\lambda = 0, \mu =
  1.0$ and (b) $\lambda = 0.1, \mu = 0$ and $\phi = 0.2\pi$. Note
  there is no current reversal when the potential is symmetric.}
\label{lam=0}
\end{figure}

\begin{figure}
  \protect\centerline{\epsfysize=3.0in \epsfbox{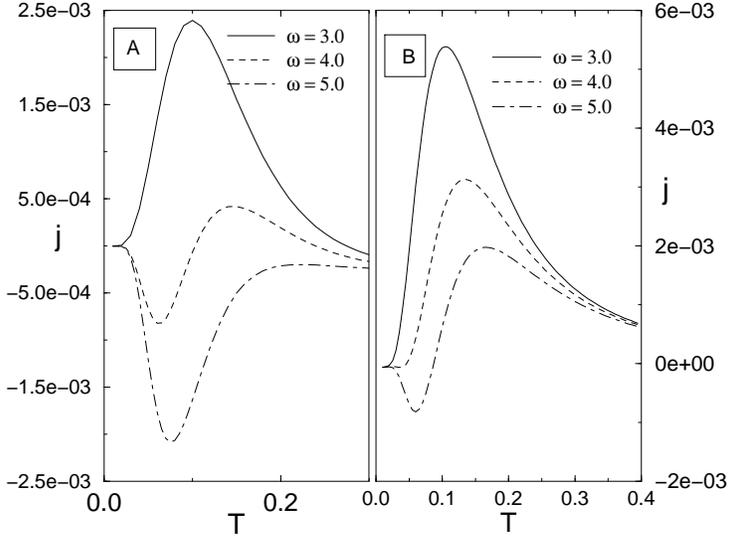}}
  \caption{ The mean current $j$ \textit{vs} temperature $T$ for $\phi =
    0.2 \pi$, $A=0.5$ and $\lambda = 0.1$. The driving frequencies are $\omega=
    3.0, 4.0$ and $5.0$. The right hand side figure shows  current $j$ 
    \textit{vs} $T$ for $\phi = 1.2 \pi$.}
  \label{j-T}
\end{figure}

                    In the Fig.~(\ref{lam=0}A), the average current $j$
is plotted as a function of temperature $T$ for different values of
$\omega$. Here the asymmetry parameter $\mu =
1.0$ and $\lambda = 0.0$.
       In this case the current reverses its sign
(only \textit{once}) for frequencies sufficiently
large as shown in the $\omega = 4.0, \omega = 5.0$ case. In the
absence of asymmetric potential and presence of space dependent
friction ( $\lambda = 1.0$), there is no current reversal irrespective
of $\omega$ and phase shift $\phi$ as shown in Fig~(\ref{lam=0}B). Hence asymmetric potential 
is essential for current reversal.However the direction of the current 
depends on the phase lag $\phi$. Separately in both these cases
 absolute value of current
 exhibits a maxima as a function of $T$, reminiscent of stochastic
 resonance phenomena. In a purely asymmetric case ($\lambda = 0$) current
 vanishes rapidly when $T$ exceeds the temperature associated with the
 barrier height. Whereas, in the symmetric case due to space dependent friction
 absolute values of currents are significantly higher and decay slowly to
 zero in the large temperature regime. Naturally in the presence of
 both asymmetry and space dependent friction for the case under study
 the low temperature regime is dominated by the effect of asymmetry while the
 high temperature regime is dominated by space dependent friction. From this
 , one can qualitatively explain the current reversals from positive
 to negative side as a function of temperature even in the adiabatic
 limit. Hence in the presence of both space
dependent friction  and asymmetric potential we can have
current reversal with $T$ even in the \textit{adiabatic regime} for a suitablily
chosen value of $\phi$ as shown in Fig~(\ref{j-T}A), where we have
plotted $j$ \textit{vs} $T$ with $\mu = 1.0$, $\lambda = 1.0$ and $\phi =
0.2 \pi$ for different values of $\omega$. 
In the presence
of finite frequency drive there are twice current reversals as shown
in the figure. This phenomena of twice current reversal with
 temperature $T$ is the foremost feature of our system, previously
 unseen in any overdamped system. When
the phase difference $\phi$ is such that the current due to space
dependent friction alone is in the same direction as that of current due to
potential asymmetry only, then we do not have current reversal in the
adiabatic case as shown in Fig~(\ref{j-T}B) where $\phi = 1.2 \pi$,
though a \textit{single} current reversal due to
finite frequency drive may be present. In all the cases studied so far 
we have observed that the current reversal don not take place above a
critical frequency $\omega_{c}$ of driving, which in turn depends
on $\phi$ and other parameters in the problem. 

%

Multiple current reversals can also be seen when the amplitude ( $A$) of the
forcing term is varied in a suitable parameter regime of our system.
\begin{figure}
 \protect\centerline{\epsfysize=3.0in \epsfbox{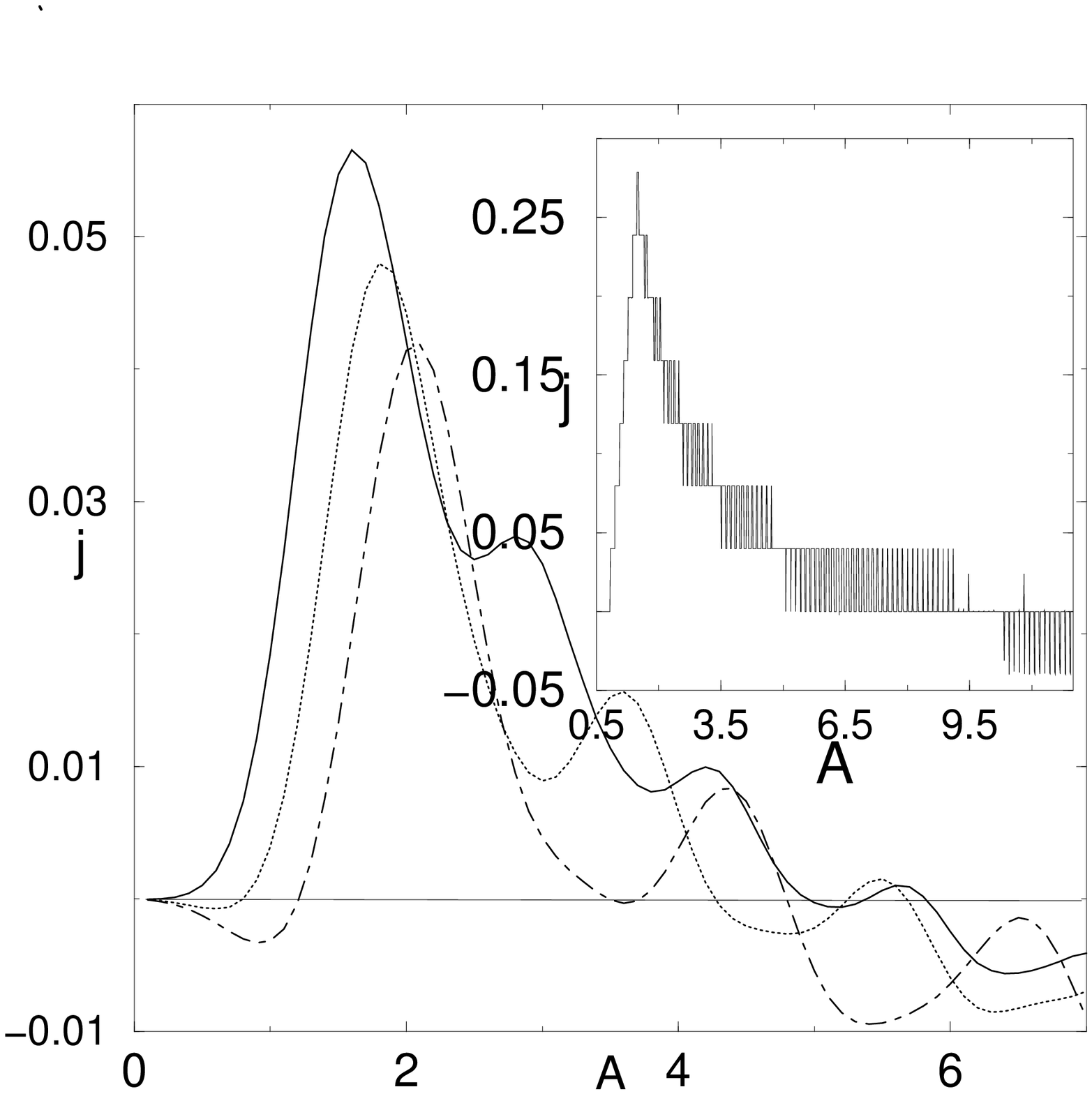}}
\caption{The mean current $j$ with amplitude $A$ of the forcing term
  for $\phi = 0.88 \pi,T=0.05$ and $\lambda=0.1$ with $\omega=3.0,
  4.0, 5.0$. The inset shows the reversal of deterministic current vs the
  amplitude of the forcing.}
\label{j-A}
\end{figure}
In Fig.~\ref{j-A}, the plot of $j$ versus $A$ is shown for
different values of $\omega$, keeping $\lambda$, $\phi$ and $T$ fixed at
0.1, $0.88\pi$ and $0.05$ respectively. For $\omega = 4.0$ curve, we can see as
many as four current reversals. For very large value of $A$, the current
asymptotically goes to a constant value depending on the value of
$\phi$, as was previously shown for the adiabatic case~\cite{Dan3}. The inset in
Fig.~\ref{j-A} shows current reversal even for the deterministic case
also. In addition it exhibits an interesting phenomena of current
quantization and phase locking. These features are washed out with
increasing noise strength. 
\begin{figure}
\protect\centerline{\epsfysize=3.0in \epsfbox{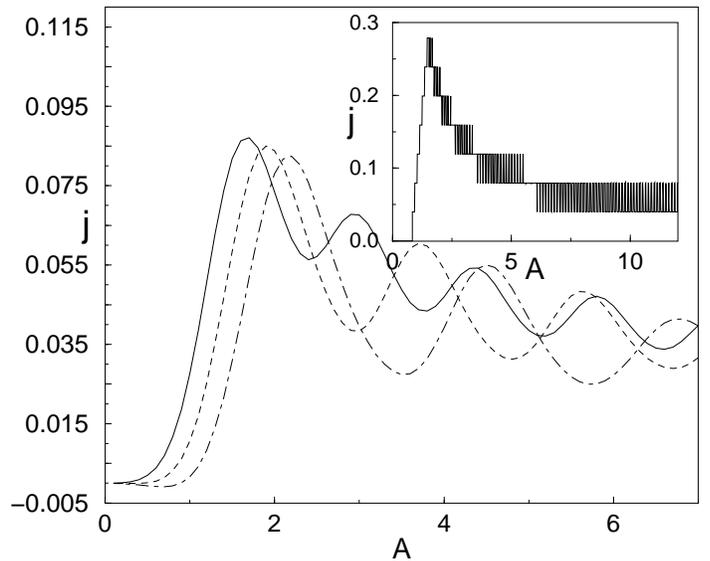}}
\caption {Mean current j vs temperature T for (a) $\lambda = 0, \mu =
  1.0$ and (b) $\lambda = 0.1, \mu = 0$ and $\phi = 0.2\pi$. Note
  there is no current reversal when the potential is symmetric.}
\label{j-A2}
\end{figure}
In Fig~(\ref{j-A2}) we have plotted  $j$ versus $A$ for $\phi =
1.2 \pi$. There is no current reversal in the adiabatic regime for
both deterministic as well as finite temperature case. The observation 
of multiple current reversals can be attributed to a cooperative interplay 
between the spatial asymmetry of the potential, the friction
inhomogeneity and the finite frequency drive. Depending on the system
parameters we may have multiple current reversal or no current
reversal at all ( see Fig.(~\ref{j-A} and \ref{j-A2})).

In conclusion, we have studied the transport properties of
overdamped Brownian particles moving in an asymmetric potential
with space dependent friction coefficient and rocked by periodic
force. We observe multiple current reversal with both temperature $T$ 
and $A$ the amplitude of the rocking force in the presence of finite
frequency driving. Current reversal also occurs in the deterministic adiabatic 
regime for suitable values of phase difference $\phi$. All the above
results can be understood qualitatively. The space dependent friction
plays an active role in these systems in 
determining several counter intuitive phenomena~\cite{Dan3,PLA,PRE}
observed in the transport processes.

\label{lastpage}

\end{multicols}
\end{document}